\documentclass[11pt]{article}

\let\fn\footnote
\renewcommand{\footnote}[1]{\linespread{1.1}\fn{#1}\linespread{1.29}}

\makeatletter\renewcommand{\section}{\@startsection
{section}{1}{\z@}{-3.5ex plus -1ex minus
    -.2ex}{2.3ex plus .2ex}{\bf }}
\makeatletter\renewcommand{\subsection}{\@startsection{subsection}{2}{\z@}{-3.25ex
plus -1ex minus
   -.2ex}{1.5ex plus .2ex}{\it }}
\makeatletter\renewcommand{\subsubsection}{\@startsection{subsubsection}{3}{-2.45ex}{-3.25ex
plus -1ex minus -.2ex}{1.5ex plus .2ex}{\it }}
\renewcommand{\thesection}{\arabic{section}.}
\renewcommand{\thesubsection}{\arabic{section}.\arabic{subsection}.}

\renewcommand{\theequation}{\thesection\arabic{equation}}
\makeatletter \@addtoreset{equation}{section}

\renewenvironment{thebibliography}[1]
     {\baselineskip=16pt plus 2pt minus 1pt
      \section*{\large\refname
        \@mkboth{\MakeUppercase\refname}{\MakeUppercase\refname}}%
     \list{\@biblabel{\@arabic\c@enumiv}}%
           {\settowidth\labelwidth{\@biblabel{#1}}%
            \leftmargin\labelwidth
            \advance\leftmargin\labelsep
            \@openbib@code
            \usecounter{enumiv}%
            \let\p@enumiv\@empty
            \renewcommand\theenumiv{\@arabic\c@enumiv}}%
      \sloppy
      \clubpenalty4000
      \@clubpenalty \clubpenalty
      \widowpenalty4000%
      \sfcode`\.\@m}

\newcommand{\acknowledgements}{\section*{Acknowledgements}
\addcontentsline{toc}{section}{\hspace{0.6cm}{\bf Acknowledgements}}}

\newcommand{\appendices}{\section*{Appendix}\setcounter{subsection}{0}\setcounter{equation}{0}\renewcommand{\thesubsection}{\Alph{subsection}.}
\renewcommand{\theequation}{\thesubsection\arabic{equation}}
\addtocontents{toc}{\vspace{0.2cm}

{\bf Appendices}}
}

\usepackage{epsfig,rotating}
\usepackage{amsmath,amssymb}
\usepackage{amsfonts}
\usepackage{mathrsfs}
\usepackage{bbm}
\usepackage{bm}

\def\slasha#1{\setbox0=\hbox{$#1$}#1\hskip-\wd0\hbox to\wd0{\hss\sl/\/\hss}}

\def\periodb#1{\setbox0=\hbox{$#1$}#1\hskip-\wd0\hbox to\wd0{-}}

\newcommand{\unit}{\mathbbm{1}}   			

\newcommand{\CA}{\mathcal{A}}    			

\newcommand{\CB}{\mathcal{B}}

\newcommand{\CC}{\mathcal{C}}

\newcommand{\CCL}{\mathscr{L}}

\newcommand{\CI}{\mathcal{I}}

\newcommand{\CL}{\mathcal{L}}

\newcommand{\CN}{\mathcal{N}}

\newcommand{\CW}{\mathcal{W}}

\newcommand{\FR}{\mathbbm{R}}     			
\newcommand{\FC}{\mathbbm{C}}     			
\newcommand{\NN}{\mathbbm{N}}     			
\newcommand{\RZ}{\mathbbm{Z}}     			

\newcommand{\dd}{\mathrm{d}}     			
\newcommand{\dpar}{\partial}     			
\newcommand{\diag}{{\mathrm{diag}}}     		
\newcommand{\di}{\mathrm{i}}     			
\newcommand{\eps}{{\varepsilon}}			

\newcommand{\bpsi}{{\bar{\psi}}}
\newcommand{\bth}{{\bar{\theta}}}
\newcommand{\bphi}{{\bar{\phi}}}
\newcommand{\bl}{{\bar{\lambda}}}

\newcommand{\bD}{{\bar{D}}}

\newcommand{\by}{{\bar{y}}}

\newcommand{\ald}{{\dot{\alpha}}}     			

\newcommand{\eand}{{~~~\mbox{and}~~~}}     		

\newcommand{\ccdot}{{\,\cdot\,}}
\newcommand{\tr}{\,\mathrm{tr}\,}     			

\newcommand{\au}{\mathfrak{u}}
\newcommand{\asu}{\mathfrak{su}}
\newcommand{\aso}{\mathfrak{so}}
\newcommand{\aspin}{\mathfrak{spin}}

\newcommand{\sU}{\mathsf{U}}     			

\newcommand{\sSU}{\mathsf{SU}}
\newcommand{\sSL}{\mathsf{SL}}

\newcommand{\sSO}{\mathsf{SO}}

\newcommand{\acton}{\vartriangleright}     			
\newcommand{\remark}[1]{}     				
\def\tyng(#1){\hbox{\tiny$\yng(#1)$}}			
\def\tyoung(#1){\hbox{\tiny$\young(#1)$}}			

\begin{document}
\begin{titlepage}
\begin{flushright}
  TCDMATH 08-10 \\
  HMI-08-02
\end{flushright}
\vskip 2.0cm
\begin{center}
{\LARGE \bf Multiple M2-branes and Generalized 3-Lie algebras}
\vskip 1.5cm
{\Large Sergey Cherkis and Christian S{\"a}mann}
\setcounter{footnote}{0}
\renewcommand{\thefootnote}{\arabic{thefootnote}}
\vskip 1cm
{\em Hamilton Mathematics Institute and\\
School of Mathematics,\\
Trinity College, Dublin 2, Ireland}\\[5mm]
{E-mail: {\ttfamily cherkis, saemann@maths.tcd.ie}} \vskip
1.1cm
\end{center}
\vskip 1.0cm
\begin{center}
{\bf Abstract}
\end{center}
\begin{quote}
We propose a generalization of the Bagger-Lambert-Gustavsson action as a candidate for the description of an arbitrary number of M2-branes. The action is 
formulated in terms of $\CN=2$ superfields in three dimensions and 
corresponds to an extension of the usual superfield formulation of 
Chern-Simons matter theories. Demanding gauge invariance of the resulting 
theory does not imply the total antisymmetry of the underlying 3-Lie algebra structure 
constants. We relax this condition and propose a class of examples for these 
generalized 3-Lie algebras. We also discuss various associated ordinary Lie algebras.
\end{quote}
\vspace{3cm}
\end{titlepage}

\section{Introduction}

Inspired by the results of \cite{Basu:2004ed}, Bagger and Lambert \cite{Bagger:2006sk,Bagger:2007jr,Bagger:2007vi} and Gustavsson \cite{Gustavsson:2007vu,Gustavsson:2008dy} constructed a theory which they conjectured to describe stacks of M2-branes, analogously to maximally supersymmetric Yang-Mills (SYM) theory describing the low-energy effective theory of multiple D-branes. The theory is an $\CN=8$ supersymmetric Chern-Simons matter theory living on a three dimensional Minkowski space. The no-go theorem for the construction of such theories \cite{Schwarz:2004yj} is circumvented by replacing the gauge algebra structure by a so-called 3-Lie algebra. Later on, it has been shown \cite{Mukhi:2008ux} that there exists a procedure\footnote{See also \cite{Distler:2008mk,Banerjee:2008pd} for an interpretation of this mechanism.} which can be interpreted as compactifying a transverse direction on a circle and which reduces the Bagger-Lambert-Gustavsson (BLG) theory to the corresponding action on multiple D2-branes plus corrections in $\frac{1}{g_{\mathrm{YM}}}$; this confirmed the original interpretation of BLG. 

Since its construction, this model has received a great deal of attention. A serious shortcoming was however encountered soon: the only 3-Lie algebra with positive definite metric which reduces to super Yang-Mills theory with gauge group $\sU(N)$ is -- in the classification of \cite{Filippov:1985aa} -- $A_4$, as was shown in \cite{Nagy:2007aa}; see also \cite{Papadopoulos:2008sk,Gauntlett:2008uf,Ho:2008bn,Papadopoulos:2008gh}. The BLG theory with this 3-Lie algebra describes two M2-branes according to the interpretation of \cite{Mukhi:2008ux}. Thus, it seems that the structure of a 3-Lie algebra has to be generalized to accommodate stacks of more than two M2-branes. Various generalizations have been proposed in the literature and we discuss them in section 2.3. The most prominent modification introduces ghosts into the theory and after removing them, one is left with the ordinary $\CN=8$ SYM theory on $\FR^{1,2}$ \cite{Bandres:2008kj}, which is not the M2-brane theory one would expect.

In this paper, we try to formulate a BLG-like theory using the $\CN=2$ superspace extension of $\FR^{1,2}$; for previous work in a similar direction see \cite{Benna:2008zy} and also \cite{Mauri:2008ai}. Our goal is to write down a fully supersymmetric theory which makes use of a triple bracket $[\ccdot,\ccdot,\ccdot]:\CA\times\CA\times\CA\rightarrow \CA$ on a vector space $\CA.$ In contrast to the BLG theory, we do not demand this bracket to be totally antisymmetric. However, we demand that the kinetic term for the gauge potential be of the same Chern-Simons form as in the BLG theory. We then impose the minimal constraints on the triple bracket to achieve gauge invariance of our theory. The result is indeed a generalization of the concept of a metric 3-Lie algebra.

We briefly review 3-Lie algebras and the BLG theory in section 2 before we write down a new superfield action in section 3. In this section, we also derive the component action and its equations of motion, and discuss some of the model's properties. Section 4 deals with generalized 3-Lie algebras: After giving the definition, we present a class of examples and discuss possible reduction mechanisms to ordinary Lie algebras.  We conclude with section 5.

\section{The BLG theory}

First, we will briefly recall the definition of metric $n$-Lie algebras as introduced by Filippov in \cite{Filippov:1985aa}, of which ordinary Lie algebras ($n=2$) and the 3-Lie algebras appearing in the BLG theory are special cases. We then review the Bagger-Lambert-Gustavsson (BLG) theory, which was essentially developed in the papers \cite{Bagger:2006sk,Bagger:2007jr,Bagger:2007vi} and \cite{Gustavsson:2007vu,Gustavsson:2008dy}. Also, we summarize some drawbacks of the conventional formulations of the theory. These motivate us to introduce a generalized 3-Lie algebra and a gauge theory Lagrangian based on it.

\subsection{Reminder: $n$-Lie algebras}

\paragraph{Definition.} Given a $\FC$-module $\CA$, define a (complex) {\em $n$-Lie algebra} \cite{Filippov:1985aa} as an algebra with an $n$-ary map $[\ccdot,\ldots ,\ccdot]:\CA^n\rightarrow \CA$ such that:

(a) $[\ccdot,\ldots ,\ccdot]$ is totally antisymmetric, i.e.\
\begin{equation}\label{Eq:antisym}
 [\tau_1,\ldots ,\tau_n]\ =\ (-1)^{|\sigma|}[\tau_{\sigma(1)},\ldots ,\tau_{\sigma(n)}]~,~~~\tau_i\in\CA
\end{equation}

(b) any $(n-1)$-plet acts via $[\ccdot,\ldots ,\ccdot]$ as a derivation, i.e.\ the bracket satisfies the {\em fundamental identity} for all $\tau_i,\rho_i\in\CA$
\begin{equation}\label{fundamentalidentityn}
 [\tau_1,\ldots ,\tau_{n-1},[\rho_1,\ldots ,\rho_n]]\ =\ \sum_{i=1}^n
[\rho_1,\ldots ,\rho_{i-1},[\tau_1,\ldots ,\tau_{n-1},\rho_i],\rho_{i+1},\ldots ,\rho_n]~.
\end{equation}

Simple examples are given by the $n$-Lie algebras \cite{Filippov:1985aa} $A_{n+1}$, $n\in \NN$: Consider an $n+1$-dimensional complex vector space $V$ with an orthonormal basis $(e_\mu)$ and define the $n$-ary product $[x_1,\ldots ,x_n]$, $x_\mu\in V$ as the determinant of the matrix $(x_1~\ldots ~x_n~e)$. In particular, the algebra $A_4$ has attracted much attention recently in the following form: Let $\CA$ be spanned by the four-dimensional $\gamma$-matrices\footnote{See the appendix for more details on our conventions.} $(\gamma^\mu)$ and let $\gamma_5=\gamma^1\ldots \gamma^4$. One can define a triple product \cite{Gustavsson:2007vu}
\begin{equation}
 [a,b,c]\ :=\ [[a,b]\gamma_5,c]~~,~~~a,b,c\in\CA~~,
\end{equation}
which makes $\CA$ into a 3-Lie algebra isomorphic to $A_4$, as one readily checks.

For simplicity, let us now restrict to the case of 3-Lie algebras, although most of the notions readily generalize to arbitrary $n$-Lie algebras. A submodule $\CI\subset \CA$ is called an {\em ideal}, if for all $i\in \CI, a,b\in \CA$, the product $[i,a,b]$ is again in $\CI$. A 3-Lie algebra $\CA$ is called {\em simple}, if $\CA\neq 0$ and $\CA$ contains no ideals except for $0$ and $\CA$. It is easy to see that the 3-Lie algebra $A_4$ is simple.

Let us now add more structure to the 3-Lie algebra. If $\CA$ as a vector space is spanned by the basis $(\tau^a)$, then the triple product is completely encoded in the structure constants $f^{abc}{}_d$ defined via
\begin{equation}
 [\tau^a,\tau^b,\tau^c]\ =:\ f^{abc}{}_d\tau^d~,
\end{equation}
and they are antisymmetric in the first three indices by definition, cf.\ \eqref{Eq:antisym}. Defining a Hermitian structure via the pairing $(\ccdot,\ccdot):\CA^2\rightarrow \FC$, we have a Hermitian matrix $h^{ab}$ from
\begin{equation}
 (\tau^a,\tau^b)\ =:\ h^{ab}~.
\end{equation}
We can use this tensor to lift indices: $f^{abce}=h^{ed}f^{abc}{}_d$. If we require that the pairing is invariant under the transformations generated by the 3-bracket, i.e.\
\begin{equation}\label{metriccompatibility}
 ([\rho^c,\rho^d,\tau^a],\tau^b)+(\tau^a,[\rho^c, \rho^d,\tau^b])\ =\ 0
\end{equation}
for all $\rho^a,\tau^a\in\CA$, then $f^{abcd}$ is totally antisymmetric, see e.g.\ \cite{Ho:2008bn}. In terms of the structure constants, the fundamental identity \eqref{fundamentalidentityn} reads
\begin{equation}\label{fundamentalidentitystruc}
 f^{efg}{}_df^{abc}{}_g\ =\ f^{efa}{}_gf^{gbc}{}_d+f^{efb}{}_gf^{agc}{}_d+f^{efc}{}_gf^{abg}{}_d
\end{equation}
and is related to contracted Pl{\"u}cker equations \cite{Ho:2008bn,Papadopoulos:2008sk}. Note that the structure constants of $A_4$ are given by $f^{abcd}=4\eps^{abcd}$.

The representations of the $n$-Lie algebras $A_n$ have been studied in \cite{Dzhuma:2002aa}; further remarks on representation theory of 3-Lie algebras are found in \cite{Kasymov1983:aa}.

\subsection{The Bagger-Lambert-Gustavsson theory}

Consider a real 3-Lie algebra $\CA$ of dimension $k$ with generators $\tau^a$, $a=1,\ldots ,k$, structure constants $f^{abc}{}_d$ and a symmetric bilinear pairing $(\ccdot,\ccdot):\CA\times\CA\rightarrow\FR$ satisfying \eqref{metriccompatibility} and giving rise to a positive definite metric $h^{ab}=(\tau^a,\tau^b)$. The field content of the BLG theory is given by eight scalar fields $X^{Ia}$, $I=1,\ldots ,8$, transforming in the vector representation of $\sSO(8)$ and world-volume Majorana spinors $\Psi^a$ having (suppressed) spinor indices of $\sSO(1,2)$ and $\sSO(8)$. Both take values in a 3-Lie algebra, as indicated by the index $a$. We work with the $32\times 32$-dimensional, anticommuting gamma matrices $(\Gamma^\mu,\Gamma^I)$, where we split the eleven-dimensional index into $(\mu,I)$ according to the branching $\sSO(1,10)\supset\sSO(1,2)\times\sSO(8)$. The spinors are the Goldstinos of this symmetry breaking and thus satisfy $\Gamma_{012}\psi^a=-\psi^a$. From the Majorana property, we obtain $\bar{\psi}^a=(\psi^a)^T \CC$, where $\CC$ is the charge conjugation operator as described e.g.\ in appendix B of \cite{Polchinski:1998rr}. In addition, there is a gauge potential $A_{\mu\,ab}=-A_{\mu\,ba}$, yielding a covariant derivative
\begin{equation}
 (\nabla_\mu X)_a\ =\ \dpar_\mu X_a-X_b\tilde{A}^b{}_a~,~~~\tilde{A}_{\mu}{}^b{}_a\ :=\ A_{\mu\,cd}f^{cdb}{}_a~.
\end{equation}
For future use, we associate a field strength $\tilde{F}_{\mu\nu}{}^a{}_b$ to this potential:
\begin{equation}\label{fieldstrength}
 \tilde{F}_{\mu\nu}{}^a{}_b\ =\ \dpar_\mu\tilde{A}_\nu{}^a{}_b-\dpar_\nu\tilde{A}_\mu{}^a{}_b+\tilde{A}_\mu{}^a{}_c\tilde{A}_\nu{}^c{}_b-\tilde{A}_\nu{}^a{}_c\tilde{A}_\mu{}^c{}_b~.
\end{equation}
The Lagrangian of the BLG theory takes the form \cite{Bagger:2007jr}
\begin{equation}\label{BLG-theory}
\begin{aligned}
 \CL_{\mathrm{BLG}}\ =\ &-\tfrac{1}{2}(\nabla_\mu X^{Ia}\nabla^\mu X^I_a)+\tfrac{\di}{2}\bar{\Psi}^a\Gamma^\mu \nabla_\mu\Psi_a\\
&+f^{abcd}\tfrac{\di}{4}\bar{\Psi}_b\Gamma_{IJ}X^I_cX^J_d\Psi_a-\tfrac{1}{12}f^{abcd}f^{efg}{}_dX_a^IX_b^JX_c^KX_e^IX_f^JX_g^K\\
&+\tfrac{1}{2}\eps^{\mu\nu\kappa}\left(f^{abcd}A_{\mu\,ab} \dpar_\nu A_{\kappa\,cd}+\tfrac{2}{3}f^{cda}{}_gf^{efgb}A_{\mu\,ab}A_{\nu\,cd}A_{\kappa\,ef}\right)~.
\end{aligned} 
\end{equation}
We chose to give the Lagrangian in terms of the structure constants rather than the 3-bracket as in this way, we avoid any ambiguities in the treatment of the fermionic fields. The equations of motion to \eqref{BLG-theory}, which had been independently derived in \cite{Gustavsson:2007vu}, are given by \cite{Bagger:2007jr} 
\begin{equation}
 \begin{aligned}
  \nabla_\mu \nabla^\mu X_a^K-\tfrac{\di}{2}\bar{\Psi}_c \Gamma^K{}_IX^I_d\Psi_b f^{bcd}{}_a+\tfrac{1}{2}f^{bcd}{}_g f^{efg}{}_a X^I_e X^J_f X^I_b X^J_c X^K_d&\ =\ 0~,\\
 \Gamma^\mu \nabla_\mu \Psi_a+\tfrac{1}{2}f^{cdb}{}_a \Gamma_{IJ}X^I_c X^J_d \Psi_b&\ =\ 0~,\\
\tilde{F}_{\mu\nu}{}^b{}_a+\eps_{\mu\nu\kappa}(X_c^J\nabla^\kappa X_d^J+\tfrac{\di}{2}\bar{\Psi}_c\Gamma^\kappa \Psi_d)f^{cdb}{}_a&\ =\ 0~.
 \end{aligned}
\end{equation}

The action arising from the Lagrangian \eqref{BLG-theory} is invariant under the supersymmetries generated by 
\begin{equation}
 \begin{aligned}
  \delta X^I_a &\ =\  \di \bar{\eps}\Gamma^I\Psi_a~,\\
  \delta \Psi_a&\ =\ \nabla_\mu X^I_a\Gamma^\mu\Gamma^I\eps-\tfrac{1}{6} X_b^IX_c^JX_d^Kf^{bcd}{}_a\Gamma^{IJK}\eps~,\\
  \delta \tilde{A}_\mu{}^b{}_a&\ =\ \di\bar{\eps}\Gamma_\mu\Gamma_I X^I_c \Psi_d f^{cdb}{}_a~,
 \end{aligned}
\end{equation}
up to equations of motion; the corresponding supersymmetry algebra closes only on shell and requiring closure of this algebra is how the equations of motion were found in the first place. Here, $\eps$ is a Majorana spinor corresponding to the 16 unbroken supersymmetries under the branching $\sSO(1,2)\times \sSO(8)\subset\sSO(1,10)$ and thus satisfies $\Gamma_{012}\eps=\eps$.

The action is simultaneously invariant under the gauge transformations of the form
\begin{equation}
 \begin{aligned}
   \delta X_d&\ =\ \tilde{\lambda}^c{}_d X_c~,~~~\delta X\ =\ \lambda_{ab}[\tau^a,\tau^b,X]~,\\
   \delta \Psi_d&\ =\ \tilde{\lambda}^c{}_d \Psi_c~,~~~\delta \Psi\ =\ \lambda_{ab}[\tau^a,\tau^b,\Psi]~,\\
   \delta A_{\mu\, ab}&\ =\ \dpar_\mu\lambda_{ab}+\tilde{\lambda}^c{}_aA_{\mu\,cb}+\tilde{\lambda}^c{}_bA_{\mu\,ac}~,\\
   \delta \tilde{A}_\mu{}^a{}_b&\ =\ \nabla_\mu \tilde{\lambda}^a{}_b=:\dpar_\mu\tilde{\lambda}^a{}_b+\tilde{A}_\mu{}^a{}_c\tilde{\lambda}^c{}_b-\tilde{\lambda}^a{}_c\tilde{A}_\mu{}^c{}_b~,\\
   \delta \tilde{F}_{\mu\nu}{}^a{}_b&\ =\ -\tilde{\lambda}^a{}_c\tilde{F}_{\mu\nu}{}^c{}_b+\tilde{F}_{\mu\nu}{}^a{}_c\tilde{\lambda}^c{}_b~,\\
 \end{aligned} 
\end{equation}
where tilded objects are defined as before, i.e.\ for example $\tilde{\lambda}^c{}_d=f^{abc}{}_d \lambda_{ab}$. As one easily verifies, the gauge algebra closes. Note that in \cite{Bagger:2007jr}, the field $\tilde{A}_\mu{}^a{}_b$ was considered physical, and only the gauge transformations for this field were given. The transformation law we gave here for the field $A_{\mu\,ab}$ is compatible with that of $\tilde{A}_\mu{}^a{}_b$ and the resulting gauge algebra closes again. 

Since we also have $\delta(\nabla_\mu X)_a=\tilde{\lambda}^c{}_a(\nabla_\mu X)_c$, gauge invariance of all terms except for the Chern-Simons term is evident. The latter is found to transform according to
\begin{equation}\label{CDtrafolaw}
 \delta \CL_{CS}\ =\ \eps^{\mu\nu\kappa}(\delta A_{\mu\,ab})~\tilde{F}_{\nu\kappa}{}^a{}_h h^{bh}~.
\end{equation}
This is the usual transformation law for Chern-Simons theory, and the gauge transformations produce a term containing a total derivative and a winding number term.

\subsection{Shortcomings of the theory}

Fixing one of the indices in the structure constants of a 3-Lie algebra $\CA$, one obtains the structure constants of the associated 2-Lie algebra $\tilde{\CA}(a_0)$ \cite{Filippov:1985aa}:
\begin{equation}
 \tilde{f}^{bc}{}_d\ =\ f^{a_0bc}{}_d~.
\end{equation}
Due to the fundamental identity \eqref{fundamentalidentitystruc}, the Jacobi identity
\begin{equation}
 \tilde{f}^{ij}{}_k\tilde{f}^{kl}{}_m+\tilde{f}^{li}{}_k\tilde{f}^{kj}{}_m+\tilde{f}^{jl}{}_k\tilde{f}^{ki}{}_m\ =\ 0
\end{equation}
is automatically satisfied. This reduction can be implemented by a Higgsing procedure \cite{Mukhi:2008ux} which effectively reduces the BLG theory to a deformed version of the Yang-Mills theory describing the low energy effective action on multiple D2 branes, cf.\ section 4.3. It has been shown in \cite{Nagy:2007aa} and later in \cite{Papadopoulos:2008sk,Gauntlett:2008uf} that essentially the only 3-Lie algebra admitting $\au(N)$ as its associated 2-Lie algebra is isomorphic\footnote{Evidently, a similar reduction process from a $n+1$-algebra to a $n$-algebra can be defined. For the series $A_n$, this reduction always ends up with $A_4$, as the structure constants are the $n$-dimensional epsilon-tensors.} to $A_4$. In this case, the Lie algebra obtained is $\asu(2)$, as $\tilde{f}^{abc}=4\eps^{abc}$. This implies that -- supposing the Higgsing procedure of \cite{Mukhi:2008ux} -- the BLG theory can only describe a stack of two M2-branes. 

To extend to a description of more than two M2-branes, there are essentially two strategies: giving up total antisymmetry of the structure constants or giving up a positive definite metric. The first approach has been followed in \cite{Gran:2008vi}, where the BLG theory has been discussed in this more general setting on the level of equations of motion. The latter approach has been followed in \cite{Gomis:2008uv}-\cite{Gomis:2008be}, see also \cite{Figueroa-O'Farrill:arXiv0805.4760}, and introduces ghosts into the theory. A detailed analysis of this situation can be found in \cite{deMedeiros:2008bf}. The results of \cite{Bandres:2008kj} seem to indicate, however, that after removing the ghosts, one arrives at maximally supersymmetric Yang-Mills theory in three dimensions and thus at exactly the low energy description of a stack of D2-branes. The absence of corrections shows that this is not the M2-brane theory one would hope for.

A third, more recently proposed variant is a reformulation of the BLG theory   \cite{Aharony:2008ug} as a $\sU(N)\times \sU(N)$ gauge theory. The latter seems particularly attractive, as it has been shown \cite{Minahan:2008hf} that this theory is integrable.

\section{Superfield formulation}

In this section, we develop a superfield formulation of BLG-like actions. This formulation automatically gives us a manifestly $\CN=2$ supersymmetric theory. The constraints we have to impose on the structure constants of the generalized 3-Lie algebra come from imposing gauge symmetry. We find that the structure constants do not have to be totally antisymmetric. Let us also stress that even with totally antisymmetric structure constants, our theory will differ from the BLG model.

\subsection{Conventions}

We follow closely the conventions of Wess and Bagger \cite{Wess:1992cp}. Three-dimensional $\CN=2$ superspace can be obtained by the usual Kaluza-Klein dimensional reduction of ordinary, four-dimensional $\CN=1$ superspace in the $x^2$-direction, i.e.\
\begin{equation}
 \sigma^{\hat{\mu}}_{\alpha\ald} \ \rightarrow\  \left\{\begin{array}{l} \sigma^{(0,1,2)}_{\alpha\ald}\ =\ \sigma^{(\hat{0},\hat{1},\hat{3})}_{(\alpha\ald)}~,~~\hat{\mu}\neq 2~,\\
\di \eps_{\alpha\ald}~,~~\hat{\mu}\ =\ 2~.
\end{array}\right.
\end{equation}
We work with signature $(-,+,+)$ and our $\eps$-conventions read
\begin{equation}
 \eps_{21}\ =\ \eps^{12}\ =\ -\eps_{12}\ =\ -\eps^{21}\ =\ 1~.
\end{equation}
The superspace $\FR^{3|4}$ has coordinates $(x^\mu,\theta^\alpha,\bth^\ald)$ with the usual reality condition that $\bth^\ald=\overline{\theta_\alpha}$. Also we use as a raising and lowering convention for spinor indices $\psi^\alpha=\eps^{\alpha\beta}\psi_\beta$ and $\psi_\alpha=\eps_{\alpha\beta}\psi^\beta$ as well as the shorthand notations $\theta^4:=\theta^2\bth^2$, $(\theta\lambda):=\theta^\alpha\lambda_\alpha$ and $(\bth\bar{\lambda}):=\bth_\ald\bar{\lambda}^\ald$. There is no longer a distinction between dotted and undotted indices in three dimensions, as all fermion fields are spinors of $\sSL(2,\FR)$ and thus real; for convenience, we nevertheless use them in our formulas. Note, however, that we defined different conventions for the contraction of either barred or unbarred spinors.

The coordinates on the chiral and anti-chiral superspaces are defined as 
\begin{equation}
 y^\mu\ :=\ x^\mu+\di \theta^\alpha\sigma^\mu_{\alpha\ald}\bth^\ald~,~~~
 \by^\mu\ :=\ x^\mu-\di \theta^\alpha\sigma^\mu_{\alpha\ald}\bth^\ald~.
\end{equation}

We use the superfield expansions with the gauge superfield in Wess-Zumino gauge as given in \cite{Wess:1992cp}. Note that bars are used instead of daggers to simplify notation:
\begin{equation}
\begin{aligned}
 \Phi^i(y)&\ =\ \phi^i(y)+\sqrt{2}(\theta \psi^i(y))+\theta^2 F^i(y)~,~~~i\ =\ 1,\ldots ,4~,\\
 \bar{\Phi}^i(\by)&\ =\ \bar{\phi}^i(\by)+\sqrt{2}(\bth \bar{\psi}^i(\by))-\bth^2 \bar{F}^i(\by)~,\\
 V_{WZ}(x)&\ =\ - \theta^\alpha\bth^\ald(\sigma^{\mu}_{\alpha\ald}A_\mu(x)+\di\eps_{\alpha\ald}\sigma(x))+\di\theta^2(\bth\bar{\lambda}(x))-\di\bth^2(\theta\lambda(x))+\tfrac{1}{2}\theta^2\bth^2 D(x)~.
\end{aligned} 
\end{equation}

As far as the 3-algebra structure is concerned, we assume that we have a real vector space $\CA$ endowed with a triple bracket $[\ccdot,\ccdot,\ccdot]:\CA\times\CA\times\CA\rightarrow\CA$ and a symmetric, bilinear, positive definite pairing $(\ccdot,\ccdot):\CA\times\CA\rightarrow \FR$. We assume that $\CA$ is finite dimensional and can be spanned by the basis $(\tau^a)$, which defines the structure constants and the metric tensor as before:
\begin{equation}
 [\tau^a,\tau^b,\tau^c]\ =\ f^{abc}{}_d\tau^d~,~~~h^{ab}\ =\ (\tau^a,\tau^b)~,~~~f^{abcd}\ =\ f^{abc}{}_e h^{ed}\ =\ ([\tau^a,\tau^b,\tau^c],\tau^d)~.
\end{equation}
No further constraints are imposed on the 3-algebra a priori.

\subsection{The superfield action}

Note that already in the paper \cite{Bagger:2006sk}, the authors gave the following superspace Lagrangian for the ungauged theory:
\begin{equation}
 \CL\ =\ c_1\int \dd^4\theta (\bar{\Phi}^i,\Phi^i)+c_2\int \dd^2\theta \eps_{ijkl}(\Phi^i,[\Phi^j,\Phi^k,\Phi^l])+c_2\int \dd^2\bth \eps_{ijkl}(\bar{\Phi}^i,[\bar{\Phi}^j,\bar{\Phi}^k,\bar{\Phi}^l])~,
\end{equation}
where $c_1,c_2$ are (real) constants. The body $\phi^i$ of the superfield $\Phi^i$ is identified with the linear combination $X^I+\di X^{I+1}$, where $I=2i-1$. In the following, we extend this superspace Lagrangian to incorporate a gauge theory. 

The $\CN=2$ superfield Lagrangian for Chern-Simons theory is well known \cite{Zupnik:1988en,Ivanov:1991fn}. This Lagrangian uses a formal parameter $t$, which is integrated over, and we fix the simplest possible $t$-dependence. For a discussion of more general choices, see \cite{Ivanov:1991fn}. Our action $S=\int \dd^3 x\CL$ is given by the Lagrangian $\CL=\CL_{CS}+\CL_{CS'}+\CL_{M}$ whose individual parts are
\begin{equation}
\begin{aligned}\label{Eq:theAction}
 \CL_{CS}&\ =\ \frac{\kappa}{2}\int_0^1\dd t\int \dd^4\theta\, V\times\left(\eps^{\alpha\ald}\bar{D}_\ald(\exp(2\di t \tilde{V}) D_\alpha(\exp(-2\di t \tilde{V})))\right)~,\\
 \CL_{CS'}&\ =\ \frac{\kappa}{2}\int_0^1\dd t\int \dd^4\theta\tfrac{1}{2}V\times\eps^{\alpha\ald}\left(-(2\di t \tilde{V})\bD_{\ald}D_{\alpha}(-2\di t\tilde{V}) +\bD_{\ald}D_{\alpha}(2\di t \tilde{V})(-2\di t\tilde{V})\right)~,\\
 \CL_{M}&\ =\ \int \dd^4\theta\,(\bar{\Phi},\Phi\exp(2\di \tilde{V}))+\int \dd^2\theta \CW(\Phi)+\int \dd^2\bth \CW(\bar{\Phi})\,~,
\end{aligned} 
\end{equation}
where the various products are defined as
\begin{equation}\label{products}
 A\times \tilde{B}\ :=\ A_{ab}~\tilde{B}^a{}_c~h^{cb}~,~~~(\tilde{A}\tilde{B})^a{}_b\ :=\ \tilde{A}^a{}_c\tilde{B}^c{}_b\eand(\Phi\tilde{A})_a\ :=\ \Phi_b\tilde{A}^b{}_a~.
\end{equation}
The superpotential $\CW(\Phi)$ is a polynomial in $\Phi$ with indices fully contracted and constructed from the triple bracket and the metric. The term $\CL_{CS'}$ is responsible for cancelling couplings between the gauge and matter fields which go beyond the minimal possible coupling. This is a new feature of the gauge theory using the triple bracket. It is trivial to see that this term would vanish if the gauge algebra were an ordinary Lie algebra\footnote{In this case, there was no distinction between $V$ and $\tilde{V}$ and a trace would enclose all terms.}. The Lagrangian is trivially supersymmetric, as all the summands are superfields and the full integral over the superspace has to transform into spacetime derivatives under supersymmetry transformations. We can thus focus on discussing the gauge invariance.  In fact, not surprisingly, all the restrictions on the structure constants of the generalized 3-Lie algebra come from imposing gauge symmetry.

First of all, we want the covariant derivative to read as 
\begin{equation}
(\nabla_\mu \phi)_a\ =\ \dpar_\mu\phi_a-\phi_c\tilde{A}^c{}_a~,
\end{equation}
and this is the reason for putting the exponential factor containing the vector superfield in the D-term on the right hand side of the scalar superfield in $\CL_M$. To reproduce the appropriate kinetic terms for scalars and fermions, we have to impose the condition
\begin{equation}\label{kintermscorrect}
h^{bc} \tilde{A}^a{}_c=:\tilde{A}^{ab}\ =\ -\tilde{A}^{ba}~~~\Leftrightarrow~~~f^{abcd}\ =\ -f^{abdc}~.
\end{equation}
This condition amounts to metric compatibility, cf.\ \eqref{metriccompatibility}, and we will come back to this point. Note that this covariant derivative can be partially integrated:
\begin{equation}
 (\nabla_\mu \phi,\psi)\ =\ -(\phi,\nabla_\mu \psi)~.
\end{equation}

We can now summarize the contributions of the various parts of the Lagrangian to the total action in terms of the component fields:
\begin{equation}
\begin{aligned}
\CL_{CS}\ =\ &\kappa\left(\eps^{\mu\nu\kappa}\left(A_\mu\times(\dpar_\nu \tilde{A}_\kappa)+\tfrac{2}{3}A_\mu\times(\tilde{A}_\nu\tilde{A}_\kappa)\right)-\di \bl_\alpha\times \tilde{\lambda}^\alpha-\di\lambda_\alpha\times\tilde{\bl}^\alpha\right.\\&\left.~~~-D\times\tilde{\sigma}-\sigma\times\tilde{D}-\tfrac{2}{3}\di A_\mu\times(\tilde{\sigma}\tilde{A}^\mu)+\tfrac{2}{3}\di A_\mu\times(\tilde{A}^\mu\tilde{\sigma})\right)~,\\
\CL_{CS'}\ =\ &\kappa\left(\tfrac{2}{3}\di A_\mu\times(\tilde{\sigma}\tilde{A}^\mu)-\tfrac{2}{3}\di A_\mu\times(\tilde{A}^\mu\tilde{\sigma})\right)~,\\
\CL_M\ =\ &(\bar{F}^i,F^i)-(\nabla_\mu \bar{\phi}^i,\nabla^\mu \phi^i)-\di(\bar{\psi}^i,\sigma^\mu\nabla_\mu \psi^i)+\di(\bar{\phi}^i,\phi^i\tilde{D})-\sqrt{2}(\bar{\phi}^i,(\psi^i\tilde{\lambda}))\\&+\sqrt{2}((\bar{\psi}^i\tilde{\bl}),\phi^i)+(\bar{\phi}^i,\phi^i\tilde{\sigma}\tilde{\sigma})-\eps^{\ald\alpha}(\bar{\psi}^i_\ald,\psi^i_\alpha\tilde{\sigma})\\
&+\int\dd^2\theta \CW(\Phi)+\int \dd^2\bth \CW(\bar{\Phi})~.
\end{aligned}
\end{equation}

\subsection{Constraints on the structure constants and gauge invariance}

There are two constraints which we have to impose on the structure constants right away: First, we demand that under the gauge symmetries $X\mapsto X+[a,b,X]$ generated by the 3-bracket, a 3-bracket of scalars should transform as a scalar. This amounts to the fundamental identity \eqref{fundamentalidentitystruc}. Second, the scalar product $(\ccdot,\ccdot)$ should be invariant under these symmetries, cf.\ \eqref{metriccompatibility}. This implies that the structure constants are antisymmetric in their last two indices, the condition we stated above in eq.~\eqref{kintermscorrect} for arriving at the desired kinetic terms for the matter fields.

One further constraint comes from gauge invariance. Consider the following gauge transformations:
\begin{equation}
\begin{aligned}
 \exp(2\di \tilde{V}')&\ =\ \exp(-\di\tilde{\Lambda})\exp(2\di \tilde{V})\exp(\di\tilde{\bar{\Lambda}})~, \\
 \Phi'&\ =\ \Phi\exp(\di \tilde{\Lambda})~,\\
\bar{\Phi}'&\ =\ \exp(-\di \tilde{\bar{\Lambda}})\bar{\Phi}~,
\end{aligned}
\end{equation}
where $\Lambda$ and $\bar{\Lambda}$ are chiral and antichiral superfields, respectively. Restricting supergauge transformations to those preserving the Wess-Zumino gauge, we obtain $\tilde{\Lambda}=\di\lambda$. Note that we work with the convention $\bar{\lambda}=-\lambda$.
We have then:
\begin{equation}
 \begin{aligned}
   \delta \phi^i_d&\ =\ \tilde{\lambda}^c{}_d \phi^i_c~,~~~\delta \phi^i\ =\ \lambda_{ab}[\tau^a,\tau^b,\phi^i]~,\\
   \delta \psi^i_d&\ =\ \tilde{\lambda}^c{}_d \psi^i_c~,~~~\delta \psi^i\ =\ \lambda_{ab}[\tau^a,\tau^b,\psi^i]~,\\
   \delta A_{\mu\,ab}&\ =\ \dpar_\mu\lambda_{ab}+\tilde{\lambda}^c{}_aA_{\mu\,cb}+\tilde{\lambda}^c{}_bA_{\mu\,ac}~,\\
   \delta \tilde{A}_\mu{}^a{}_b&\ =\ \nabla_\mu \tilde{\lambda}^a{}_b=:\dpar_\mu\tilde{\lambda}^a{}_b+\tilde{A}_\mu{}^a{}_c\tilde{\lambda}^c{}_b-\tilde{\lambda}^a{}_c\tilde{A}_\mu{}^c{}_b~,\\
 (\delta D)_{ab}&\ =\ \tilde{\lambda}^c{}_aD_{cb}+\tilde{\lambda}^c{}_bD_{ac}~,~~~\delta\sigma\ =\ \tilde{\lambda}^c{}_a\sigma_{cb}+\tilde{\lambda}^c{}_b\sigma_{ac}~.
 \end{aligned} 
\end{equation}
Closure of the gauge algebra is again immediate and does not require any constraints on the structure constants.
Gauge invariance of the action implies that the Chern-Simons term should transform according to
\begin{equation}\label{CStrafolaw2}
 \delta(\eps^{\mu\nu\kappa}\left(A_\mu\times(\dpar_\nu \tilde{A}_\kappa)+\tfrac{2}{3}A_\mu\times(\tilde{A}_\nu\tilde{A}_\kappa)\right))\ =\ \eps^{\mu\nu\kappa}(\delta A_{\mu})\times\tilde{F}_{\nu\kappa}~,
\end{equation}
cf.\ \eqref{CDtrafolaw}. Here, the field strength $\tilde{F}_{\nu\kappa}$ is defined in \eqref{fieldstrength}. For the transformation law \eqref{CStrafolaw2}, we needed the property $f^{abcd}=f^{cdab}$ of the structure constants. Altogether, the total symmetry properties of the structure constants are
\begin{equation}\label{symprop1}
 f^{abcd}\ =\ -f^{bacd}\ =\ -f^{abdc}\ =\ f^{cdab}~.
\end{equation}
Using these relations and the fundamental identity, we can easily check the gauge invariance of the action, e.g.\
\begin{equation}
\begin{aligned}
 \delta(D\times \tilde{\sigma})&\ =\ (\delta D)\times \tilde{\sigma}+D\times (\delta \tilde{\sigma})\\
&\ =\ D_{eb}\lambda_{gh}\sigma_{cd}\left(f^{abcd}f^{ghe}{}_a+f^{eacd}f^{ghb}{}_a+f^{ebad}f^{ghc}{}_a+f^{ebca}f^{ghd}{}_a\right)\ =\ 0~,
\end{aligned}
\end{equation}
The terms arising from the superpotential are gauge invariant, as long as all expressions are constructed from the pairing $(\ccdot,\ccdot)$ and the three-bracket $[\ccdot,\ccdot,\ccdot].$

One should stress that the symmetry properties given above in \eqref{symprop1} guarantee supersymmetry as well as gauge invariance of our action \eqref{Eq:theAction}.  These symmetry properties, however, are not sufficient to render the original BLG theory supersymmetric, while its gauge invariance is guaranteed by the fundamental identity and the metric compatibility condition.

\subsection{Component action and equations of motion}

Having fixed the symmetry properties of the structure constants $f^{abcd}$ to \eqref{symprop1}, we can integrate out the auxiliary fields $D,\sigma,F^i$ and $\lambda_\alpha$ to arrive at the component action and determine the equations of motion. The Lagrangian then reads
\begin{equation}\label{componentaction}
\begin{aligned}
\CL\ =\ &\kappa\eps^{\mu\nu\kappa}\left(A_\mu\times(\dpar_\nu \tilde{A}_\kappa)+\tfrac{2}{3}A_\mu\times(\tilde{A}_\nu\tilde{A}_\kappa)\right)-(\nabla_\mu \bar{\phi}^i,\nabla^\mu \phi^i)-\di(\bar{\psi}^i,\sigma^\mu\nabla_\mu \psi^i)\\
&+\frac{1}{4\kappa^2}\left([\bphi^i,\phi^i,\phi^j],[\bphi^k,\phi^k,\bphi^j]\right)-\frac{\di}{2\kappa}\left([\bphi^i,\phi^i,\bpsi^j_\alpha],\psi^{j\,\alpha}\right)+\frac{\di}{\kappa}\left([\bpsi^{j\,\alpha},\phi^j,\bphi^i],\psi^i_\alpha\right)\\
&+\int\dd^2\theta \CW(\Phi)+\int \dd^2\bth \CW(\bar{\Phi})~.
\end{aligned}
\end{equation}
Note that one has to fix a convention for how to treat fermionic fields and their interchange in a triple bracket. The bracket we use is independent of parity in the following sense:
\begin{equation}
 ([A,B,C],D)\ :=\ A_aB_bC_cD_d([\tau^a,\tau^b,\tau^c],\tau^d)\ =\ A_aB_bC_cD_d f^{abcd}~.
\end{equation}

Putting the superpotential terms to zero, we obtain the corresponding equations of motion:
\begin{equation}
 \begin{aligned}
	2\tilde{F}_{\rho\sigma}{}^a{}_b+\eps_{\mu\rho\sigma}\left(\di\bpsi^i_e\sigma^\mu\psi^i_f-\bphi^i_e(\nabla^\mu\phi^i)_f+(\nabla^\mu\bphi^i)_e\phi^i_f\right)f^{efa}{}_b&\ =\ 0~,\\
 	(\sigma^\mu\nabla_\mu \psi^i)_\ald-\frac{1}{2\kappa}[\bphi^j,\phi^j,\psi^i_\ald]-\frac{1}{\kappa}[\bphi^j,\psi^j_\ald,\phi^i]&\ =\ 0~,\\
	\nabla_\mu\nabla^\mu\phi^i-\frac{1}{2\kappa^2}[\phi^j,[\bphi^k,\phi^k,\bphi^j],\phi^i]+\frac{1}{4\kappa^2}[\bphi^k,\phi^k,[\bphi^j,\phi^j,\phi^i]]&\\-\frac{\di}{2\kappa}[\bpsi^j_\alpha,\psi^{j\,\alpha},\phi^i]+\frac{\di}{\kappa}[\bpsi^{j\,\alpha},\phi^j,\psi^i_\alpha]&\ =\ 0~.
 \end{aligned}
\end{equation}
Examples of the superpotential terms involving one triple bracket and yielding nontrivial contributions are
\begin{equation}\label{superpotentialterms}
 \CW_\alpha(\Phi)\ =\ \alpha\eps_{ijkl}([\Phi^i,\Phi^j,\Phi^k],\Phi^l)\eand
 \CW_\beta(\Phi)\ =\ \beta([\Phi^i,\Phi^j,\Phi^i],\Phi^j)~.
\end{equation}
The first term reproduces the potential terms of the BLG theory; the second term evidently vanishes if the triple bracket is given by that of a 3-Lie algebra.

\subsection{Reduced R-symmetry and generalizations}

Rewriting the component action \eqref{componentaction} using real scalar fields $X^I$, one notices that the symmetry group $\sSO(8)$ mixing the eight real scalar fields (as well as their fermionic superpartners) of the free action is broken by the matter field potential down to $\sU(4)$. This is reminiscent of the case of $\CN=4$ super Yang-Mills theory in four dimensions written in terms of $\CN=1$ superfields, where the full R-symmetry group was only obtained since the couplings of the superpotential terms are tuned to a specific value. Let us therefore look more carefully at the superpotential terms at hand \eqref{superpotentialterms}. The first term $\CW_\alpha(\Phi)$ breaks $\sU(4)$ down to $\sSU(4)\times{\mathbb Z}_2$ and produces the BLG interaction terms which we know to be invariant under $\sSO(8)$. (All the indices $I,J,K$ appearing after replacing $\phi^i\rightarrow X^{I}+\di  X^{I+1}$ with $I=2i-1$ are contracted with $\delta_{IJ}$.) The second term $\CW_\beta(\Phi)$, however, would break the R-symmetry group even more severely than the potential terms arising from integrating out the auxiliary fields $D$ and $\sigma$: the (manifest) resulting subgroup would be $\sSO(4)$. One easily verifies that adding arbitrary combinations of these superpotential terms to the action cannot restore the $\sSO(8)$ invariance. Without the potential terms, the Lagrangian (\ref{componentaction}) is $\sU(4)$ invariant.  Whether the proposed Lagrangian has enhanced supersymmetry and what part of this $\sU(4)$  symmetry is an R-symmetry and what part is just the flavour symmetry remains to be seen.

An unlikely solution to the problem of restoring the full R-symmetry group might be a deformation of the D-term in the action, which corresponds to assuming that the target space of the M2-branes is not flat space but has a nontrivial K{\"a}hler potential. It is perceivable that a suitable deformation yields an enlarged R-symmetry group. 

We should stress that after restricting to an ordinary 3-Lie algebra, our theory does not quite reproduce the BLG theory, but comes with additional terms in the potential, and thus the theories are necessarily different.

We have, however, quite an amount of freedom in deforming our theory. In particular, one could add a supersymmetric Yang-Mills-Higgs term. A theory containing both a topological term and a Yang-Mills terms usually has interesting duality properties and therefore one should certainly examine this deformation in more detail.

\section{Generalized 3-Lie algebras}

In this section, we formalize our findings from the previous section and introduce the notion of a generalized 3-Lie algebra. We also give a class of examples for this structure and discuss various reduction mechanisms, which allow for obtaining ordinary Lie algebras from the generalized 3-Lie algebras.

\subsection{Definition}

Our generalization of the notion of an $n$-Lie algebra essentially amounts to relaxing total antisymmetry of the structure constants:

\paragraph{Definition.} Given an $\FR$-module $\CA$, we define a real {\em generalized 3-Lie algebra with pairing} as an algebra $\CA$ with a ternary map $[\ccdot,\ccdot,\ccdot]:\CA^3\rightarrow \CA$ and a symmetric, bilinear, positive definite pairing $(\ccdot,\ccdot):\CA^2\rightarrow \FR$ satisfying the following properties:

1) {\em fundamental identity:}
\begin{equation}\label{fundamentalidentity}
 {}[x,y,[a,b,c]]\ =\ [[x,y,a],b,c]+[a,[x,y,b],c]+[a,b,[x,y,c]]
\end{equation}

2) invariance of the pairing or {\em metric compatibility condition:}
\begin{equation}\label{metriccompatibilitygen}
 {}([x,y,a],b)+(a,[x,y,b])\ =\ 0
\end{equation}

3) the additional symmetry property:
\begin{equation}
 ([x,y,a],b)\ =\ ([a,b,x],y)
\end{equation}

\noindent for all $x,y,a,b,c\in\CA$.

\

The first condition guarantees that a 3-bracket of scalars transforms as a scalar. The second property guarantees the invariance of the pairing $(\cdot,\cdot)$. The third property seems to be crucial in defining gauge invariant, supersymmetric actions as demonstrated above.

The pairing allows us to introduce a metric corresponding to a basis $(\tau^a)$ by
\begin{equation}
 h^{ab}\ =\ (\tau^a,\tau^b)~,
\end{equation}
and, since the pairing is positive definite, we can raise and lower indices using this metric. Structure constants $f^{abc}{}_d\in\FR$ are introduced as for 3-Lie algebras:
\begin{equation}
 [\tau^a,\tau^b,\tau^c]\ =\ f^{abc}{}_d\tau^d\eand f^{abcd}\ =\ f^{abc}{}_eh^{de}~.
\end{equation}
The conditions we imposed above on our generalized 3-Lie algebra with pairing can be reformulated using the structure constants. The fundamental identity reads as
\begin{equation}
 f^{efg}{}_df^{abc}{}_g\ =\ f^{efa}{}_gf^{gbc}{}_d+f^{efb}{}_gf^{agc}{}_d+f^{efc}{}_gf^{abg}{}_d~,
\end{equation}
and the remaining conditions are captured by the symmetry properties
\begin{equation}
 f^{abcd}\ =\ -f^{bacd}\ =\ -f^{abdc}\ =\ f^{cdab}~.
\end{equation}

A submodule $\CI\subset \CA$ is called a {\em left ideal}, if for all $i\in\CI,a,b\in\CA$, we have $[a,b,i]\in \CI$; it is called a {\em right ideal}, if for all $i\in\CI,a,b\in\CA$, we have $[i,a,b]\in \CI$ instead. A generalized 3-Lie algebra is called simple, if the only ideals it contains are the trivial ones.

\subsection{The generalized 3-Lie algebras $\CC_{2d}$}

Let us now present a class of examples of generalized 3-Lie algebras which are motivated by the original 3-bracket in \cite{Gustavsson:2007vu}. Consider the vector space $V$ of Hermitian matrices of dimension $2d\times 2d$ and define $\Gamma_{ch}=\diag(\unit_n,-\unit_n)\in V$. Note that the vector space $V$ splits into the direct sum $V=V_0\oplus V_1$ with 
\begin{equation}
\begin{aligned}
 \Gamma_{ch} a_0\ =\ +a_0 \Gamma_{ch}~,~~~a_0\in V_0~,\\
 \Gamma_{ch} a_1\ =\ -a_1 \Gamma_{ch}~,~~~a_1\in V_1~. 
\end{aligned}
\end{equation}
We take $\CA=V_1$ and, using the commutator $[a,b]:=ab-ba,$ define the ternary operation $[\ccdot,\ccdot,\ccdot]: \CA^3\rightarrow \CA$ as
\begin{equation}\label{l2dbracket}
 [a_1,a_2,a_3]\ \mapsto\  [[a_1,a_2]\Gamma_{ch},a_3]~,~~~a_1,a_2,a_3\in\CA~.
\end{equation}
A slightly tedious calculation shows that the fundamental identity \eqref{fundamentalidentity} is satisfied. 

As the bracket is not totally antisymmetric in general, $\CA$ satisfies only the requirements for being a generalized 3-Lie algebra. If we antisymmetrized the bracket \eqref{l2dbracket}, however, we would loose the fundamental identity. 

We can evidently define a symmetric positive definite pairing on $\CA$ by
\begin{equation}\label{l2dpairing}
 (a,b)\ :=\ \tr(ab)~,~~~a,b,\in\CA~,
\end{equation}
 which satisfies the compatibility condition \eqref{metriccompatibilitygen}:
\begin{equation}
 \tr([t_1,t_2,a]b)+\tr(a [t_1,t_2,b])\ =\  0~,
\end{equation}
as one readily verifies by a direct computation.  We denote this generalized 3-Lie algebra $\CA$ with the 3-bracket \eqref{l2dbracket} and the pairing \eqref{l2dpairing} $(\CA,[\ccdot,\ccdot,\ccdot],(\ccdot,\ccdot))$ by $\CC_{2d}$.

Note that in the case of $\CC_4$, we can restrict $V_1$ to the four-dimensional vector subspace spanned by the gamma matrices in four dimensions, upon which it turns into the 3-Lie algebra $A_4$. 

As a (real) basis for $\CA$, we can use products of odd numbers of gamma matrices
\begin{equation}\label{basis}
 \gamma^i~,~~~\di\gamma^{ijk}~,~~~\gamma^{ijklm}~,~~~\ldots ~~.
\end{equation}
In this basis, the pairing in $\CA$ reduces to the ordinary scalar product: 
\begin{equation}
 (\gamma^A,\gamma^B)\ :=\ \tr(\gamma^A\gamma^B)\ =\ 2d\, \delta^{AB}~~,
\end{equation}
where $A,B$ are ordered multi-indices. The expression $\delta^{AB}$ vanishes, unless the indices contained in $A$ are the same as the ones in $B$. Note that this Killing metric is positive definite.

We can now define the structure constants
\begin{equation}
 f^{ABCD}\ =\ \tr([[\gamma^A,\gamma^B]\Gamma_{ch},\gamma^C]\gamma^D)~,
\end{equation}
where $A,B,C,D$ are multi-indices. The symmetry properties of the structure constants are summarized in
\begin{equation}
 f^{ABCD}\ =\ -f^{BACD}~,~~f^{ABCD}\ =\ -f^{ABDC}~,~~f^{ABCD}\ =\ f^{CDAB}~.
\end{equation}

Note that with the basis \eqref{basis}, it is not difficult to see that $\CC_{2d}$ is simple: Given two elements $\gamma^A$ and $\gamma^B$ of the basis of $\CA$, where $A$ and $B$ are multi-indices, it is always possible to find basis elements $\gamma^C$ and $\gamma^D$ such that $[\gamma^A,\gamma^C,\gamma^D]=\gamma^B$.

\subsection{Comments on the Higgs mechanism}

To pass from a stack of M2-branes to a stack of D2-branes, it is necessary to compactify the target space of the theory along a transverse direction. In the paper \cite{Mukhi:2008ux}, a procedure for performing this reduction has been proposed. It is based on compactifying a transverse direction on a circle with radius $R$, which in turn is interpreted as fixing the value of the corresponding scalar field. By $\sSO(8)$-invariance, one can choose to fix
\begin{equation}
 \langle X^8\rangle \ =\  \frac{R}{\ell^{3/2}_p}\ =\ \sqrt{\frac{g_s}{\ell_s}}\ =\ g_{\mathrm{YM}}~,
\end{equation}
where $R$ is the compactification radius, $\ell_p$ the Planck length, $g_s$ and $\ell_s$ the string coupling constant and the string length, respectively, and $g_{\mathrm{YM}}$ the Yang-Mills coupling constant in the effective field theory on the D2-brane. This procedure, when applied to the BLG theory with a 3-Lie algebra given by the structure constants $f^{abcd}=\eps^{abcd}$, results -- according to \cite{Mukhi:2008ux} -- in a theory which is a deformation of $\CN=8$ SYM theory with gauge group $\sSU(2)$ in three dimensions with the deformation parameter $\frac{1}{g_{\mathrm{YM}}}$. That is, in the strong coupling limit, both theories agree. The gauge algebra arises here by fixing one index of the 3-Lie algebra structure constant: $\eps^{abc}=\eps^{abc4}$.

If the generalized 3-Lie algebra is a 3-Lie algebra, i.e.\ the structure constants are totally antisymmetric, this procedure applies also to our case with the Lagrangian \eqref{Eq:theAction}. However, it should yield a different deformation.

\subsection{Associated Lie subalgebras}

Let us now study a similar procedure in the case of $\CC_4$ in more detail. We choose again the basis \eqref{basis} and select $\tau^H=\gamma^4$ as the element with respect to which we want to reduce. Explicitly, our basis reads as
\begin{equation}
 \tau^A\ =\ (\gamma^1,\ldots ,\gamma^4,\di\gamma_5\gamma^1,\ldots ,\di\gamma_5\gamma^4)~.
\end{equation}
We have the pairing $(\tau^A,\tau^B)=4\delta^{AB}$ and the non-vanishing structure constants of the form $f^{ij41}$ read\footnote{We list only non-vanishing  components up to obvious symmetries.}
\begin{equation}
 f^{1243}\ =\ -f^{1342}\ =\ f^{2341}\ =\ -4~,~~f^{5643}\ =\ -f^{5742}\ =\ f^{6741}\ =\ -4~.
\end{equation}
Then the set of generators $\{\gamma^1,\gamma^2,\gamma^3,\di\gamma_5\gamma^4\}$ spans a Lie subalgebra with respect to the bracket
\begin{equation}
 [\tau^A,\tau^B]\ :=\ [\tau^A,\tau^B,\gamma^4]~.
\end{equation}

Another way of identifying a {\em labelled Lie subalgebra} for a given element (the label) $s$ in a generalized 3-Lie algebra is to find a maximal set of elements $\CCL_s$ such that
\begin{equation}
[s,a,b]\ =\ [a,b,s]~~~\mbox{for all}~a,b\in\CCL_s~.
\end{equation}
In this case, the fundamental identity guarantees that the Jacobi identity is satisfied for all elements $a,b\in\CCL_s$ with the Lie bracket $[a,b]:=[a,b,s]$.

\subsection{The associated Lie algebra by combination}

Let us now associate a Lie algebra to a (generalized) 3-Lie algebra in a different way. Elements of $\CA\times \CA$ define a map $\phi:\CA\mapsto \CA$
\begin{equation}
 \phi_{(a_1,a_2)}(b)\ :=\ [a_1,a_2,b]~~,~~a_1,a_2,b \in \CA~~.
\end{equation}
Note that $\phi_{(a_1,a_2)}=-\phi_{(a_2,a_1)}$ and $\phi_{(a_1,a_1)}=0$. We denote the set of all pairs in $\CA$ modulo equivalence and triviality by $\CB_{\CA}$. The commutator of two elements in $\CB_\CA$ is again in $\CB_\CA$:
\begin{equation}
\begin{aligned}
{}[(a_1,a_2),(b_1,b_2)]\acton v\ :& =\  
 \phi_{(a_1,a_2)}(\phi_{(b_1,b_2)}(v))-\phi_{(b_1,b_2)}(\phi_{(a_1,a_2)}(v)) \\
 &\ =\ [a_1,a_2,[b_1,b_2,v]]-[b_1,b_2,[a_1,a_2,v]]\\
&\ =\  [[a_1,a_2,b_1],b_2,v]+[b_1,[a_1,a_2,b_2],v]\\
&\ =\ \big(\phi_{([a_1,a_2,b_1],b_2)}+\phi_{(b_1,[a_1,a_2,b_2])}\big)(v)\\
&\ =\ -[[b_1,b_2,a_1],a_2,v]-[a_1,[b_1,b_2,a_2],v]~,
\end{aligned}
\end{equation}
where $a_1,a_2,b_1,b_2,v\in\CA$ and we used the fundamental identity in the third line. This expression is clearly contained in $\CB_\CA$. Furthermore, the Jacobi identity for the commutator $[(\ccdot,\ccdot),(\ccdot,\ccdot)]$ is satisfied, and $(\CB_\CA,[(\ccdot,\ccdot),(\ccdot,\ccdot)])$ forms a Lie algebra.

Let us briefly recall the example of the 3-Lie algebra $A_4$ and examine the associated Lie algebra arising by combination. Recall that the algebra $A_4$ is spanned by $\gamma$-matrices in four dimensions (see the appendix) and endowed with the triple product
\begin{equation}
 [A,B,C]\ :=\ [[A,B]\gamma_5,C]~.
\end{equation}
The associated Lie algebra $\CB_{A_4}$ is spanned by pairs of $\gamma$-matrices and its bracket is easily computed:
\begin{equation}
 [(\gamma_i,\gamma_j),(\gamma_k,\gamma_l)]\acton v\ =\ [\gamma^{ij},\gamma^{kl}] v~.
\end{equation}
Thus, $\CB_{A_4}$ is isomorphic to the Lie algebra $\aspin(4)\cong \aso(4)$.

The fact that the family of generalized 3-Lie algebras $\CC_{2d}$ is simple, implies that one can construct an associated Lie algebra by combination also in this case. Being simple translates here into two ``fundamental actions'' closing into a third one. That is, for any $a_1,b_1,a_2,b_2,x\in \CA$ there are constants $\lambda_{AB}$ such that , 
\begin{equation}\label{fundamentalactioncloses}
 \phi_{(a_1,b_1)}(\phi_{(a_2,b_2)}(x))\ =\ [a_1,b_1,[a_2,b_2,x]]\ =\ \lambda_{AB}[\gamma^A,\gamma^B,x]~.
\end{equation}
Let us study the case $\CC_4$ in more detail. The basis we use is again the one given in \eqref{basis}. We associate a map $\phi:\CA\mapsto\CA$ to pairs of elements $(\tau^A,\tau^B)\in\CA^2$ via
\begin{equation}
 \phi_{(\tau^A,\tau^B)}(v)\ :=\ [\tau^A,\tau^B,v]~.
\end{equation}
By definition, we have $\phi_{(\tau^A,\tau^A)}(v)=0$ and $\phi_{(\tau^A,\tau^B)}(v)=-\phi_{(\tau^B,\tau^A)}(v)$, $v\in\CA$. Further identities are 
\begin{equation}
 \phi_{(\tau^A,\tau^B)}(v)\ =\ -\phi_{(\gamma_5\tau^A,\gamma_5\tau^B)}(v)
\end{equation}
and
\begin{equation}
 \phi_{(\gamma^i,\gamma_5\gamma^j)}(v)\ =\ 0~.
\end{equation}
Thus, the nontrivial generators for the Lie algebra are $(\gamma^{[i},\gamma^{j]})=-(\gamma_5\gamma^{[i},\gamma_5\gamma^{j]})$ and taking the sum and the difference of these generators, we learn that the Lie algebra is $\aso(4)$ plus 6 generators acting trivially in the representation given by the triple bracket. We can therefore associate them with additional $\au(1)$-charges.

\subsection{Remarks on further structures involving 3-brackets}

In \cite{Gran:2008vi}, it was observed that the necessary condition for the closure of the SUSY algebra in the Bagger-Lambert theory is not the fundamental identity, but the weaker condition
\begin{equation}\label{weakfundamentalident}
 f^{[abc}{}_gf^{d]ge}{}_f\ =\ 0~,
\end{equation}
which defines a {\em weak 3-Lie algebra}. This condition is equivalent to the fundamental identity for totally antisymmetric structure constants. In other cases, it allows for a trivial lift of a Lie algebra with some structure constants $\tilde{f}^{ij}{}_k$ to a 3-Lie algebra by defining
\begin{equation}
 f^{\phi ij}{}_k\ =\ \tilde{f}^{ij}{}_k~.
\end{equation}
Such structure constants, however, cause problems in the Lagrangian formulation of the Baggert-Lambert theory and one has to work at the level of equations of motion \cite{Gran:2008vi}.

Note that a weak 3-Lie algebra does not always allow for an associated Lie algebra by combination. As an example, consider the algebra $\CA$ generated by 5 elements $\tau^a$, the metric $h^{ab}=\delta^{ab}$ and the non-vanishing (and not totally antisymmetric) structure constants
\begin{equation}
 f^{1245}\ =\ 1~,~~f^{1354}\ =\ 1~.
\end{equation}
The weak fundamental identity \eqref{weakfundamentalident} is trivially satisfied. However,
\begin{equation}
 \phi_{(\tau^1,\tau^3)}(\phi_{(\tau^1,\tau^2)}(\tau^4))-\phi_{(\tau^1,\tau^2)}(\phi_{(\tau^1,\tau^3)}(\tau^4))\ =\ \tau^4
\end{equation}
and there are no $\lambda_{ab}$ such that $\lambda_{ab}\phi_{(\tau^a,\tau^b)}(\tau^4)=\tau^4$. In other words, the commutator of the actions of $(\tau^1,\tau^3)$ and $(\tau^1,\tau^2)$ cannot be represented by another pair action.

Similarly, since the structure constants are not totally antisymmetric, it is clear that a Lie algebra by the usual reduction procedure can be constructed only for a small set of weak 3-Lie algebras.

A different class of ternary algebras is formed by the so-called {\em Lie triple systems} \cite{Lister:1952aa}. The 3-bracket $[\ccdot,\ccdot,\ccdot]$ in such a system $\CA$ satisfies (amongst others) the equation
\begin{equation}\label{condtriplesystem}
 [x,y,z]+[y,z,x]+[z,x,y]\ =\ 0~,~~x,y,z\in\CA~,
\end{equation}
Furthermore, the map $[x,y,\ccdot]:\CA\rightarrow\CA$ with $x,y\in\CA$ acts again as a derivation, i.e.\ it satisfies the fundamental identity. It is evident that a 3-Lie algebra with a totally antisymmetric 3-bracket satisfying \eqref{condtriplesystem} is trivial. Contrary to an ordinary 3-Lie algebra, a {\em generalized 3-Lie algebra} can in fact also be a Lie triple system, as the total antisymmetry yielding triviality is no longer present.

The classification of such Lie triple systems relies on embedding them into a $\RZ_2$-graded algebra \cite{Lister:1952aa,Faulkner:1980aa}; for more recent work see e.g.\ \cite{Benito:1999aa}.

\section{Discussion and outlook}

In this paper, we presented a Lagrangian of a new supersymmetric gauge theory which might be relevant to the description of a stack of multiple M2-branes. The theory was formulated using superfields and it has the same Lagrangian as the BLG theory up to additional potential terms for the matter fields. Demanding gauge invariance of the action imposed certain conditions on the involved structure constants $f^{abcd}$, which led us to the concept of a generalized 3-Lie algebra. We gave a class of examples for such a generalized 3-Lie algebra and identified associated ordinary Lie algebras. 

If we do not include the superpotential terms, the interaction terms for the matter fields (which differ from the BLG theory) still break the R-symmetry group from the original $\sSO(8)$ of the BLG theory down to a subgroup. This fact might only be curable, however unlikely, by changing the K{\"a}hler potential of the target space, which is a rather drastic step. Although this feature is clearly a disadvantage of our theory compared to the BLG model, studying the action proposed in this paper might nevertheless tell us much about the uniqueness of supersymmetric 3-Lie algebra gauge theories.

Needless to say that there arise several directions for further research from our discussion. The first one concerns a detailed study of the extended supersymmetry and conformal invariance of the theory. The second direction is to examine the various deformations of our theory (as e.g.\ adding a Yang-Mills-Higgs term and choosing a nontrivial K{\"a}hler potential). A third point is to develop a general classification of generalized 3-Lie algebras, as it has been done for ordinary 3-Lie algebras and Lie triple systems, as well as to study the associated Lie algebra structures. The question of reduction and the Higgs mechanism is intimately related to this point. The fourth direction is certainly to study the quantum properties of our theory, as e.g.\ done in \cite{Gustavsson:2008bf} for the BLG theory. This should be facilitated by having a superfield formulation at hand. In particular, it would be interesting to extract the restrictions on the choice of superpotential $\CW$ imposed by demanding renormalizability. Eventually, there is also the very important question of integrability of the theory. Recall that the $\sU(N)\times\sU(N)$ gauge theory recently proposed in \cite{Aharony:2008ug} has been shown to come with a dilatation operator linked to integrable spin chains \cite{Minahan:2008hf}. A similar result for our theory would certainly be most desirable. 

\paragraph{Note added.} While this manuscript was being finished, the paper \cite{Bagger:2008se} appeared, in which the total antisymmetry of the 3-bracket was also relaxed.

\acknowledgements
We would like to thank Brian Dolan for pointing out the possible consequences of an additional Yang-Mills-Higgs term in the action and Masahito Yamazaki for comments on the first version of this paper. SCh is supported by the Science Foundation Ireland Grant No.\ 06/RFP/MAT050 and by the European Commission FP6 program MRTN-CT-2004-005104. CS is supported by an IRCSET (Irish Research Council for Science, Engineering and Technology) postdoctoral fellowship.

\appendices

\subsection{$\gamma$-matrices and Clifford algebras}

We use the following conventions for the gamma matrices generating the Clifford algebras in various dimensions with Euclidean signature:
\begin{equation}
 \{\gamma^\mu,\gamma^\nu\}\ =\ 2\delta^{\mu\nu}~,~~~(\gamma^\mu)^\dagger\ =\ \gamma^\mu~.
\end{equation}
In four dimensions, we work with the following explicit set:
\begin{equation}
 \gamma^\mu\ :=\ \left(\begin{array}{cc}
        0 & \sigma^\mu \\ \bar{\sigma}^\mu & 0
       \end{array}\right)~,~~~\gamma_5\ :=\ \gamma^1\gamma^2\gamma^3\gamma^4\ =\ \left(\begin{array}{cc}
        \unit & 0 \\ 0 & -\unit
       \end{array}\right)~,
\end{equation}
where $\sigma^\mu:=(-\di\vec{\sigma},\unit)$, $\bar{\sigma}^\mu:=(\di \vec{\sigma},\unit)$. Note that we use these conventions only in the definition of the generalized 3-Lie algebras $\CC_{2d}$, while on three-dimensional superspace $\FR^{3|8}$, we followed the convention of Wess and Bagger \cite{Wess:1992cp}. 

With our conventions in four dimensions, we have $(\gamma^\mu)^\dagger=\gamma^\mu$ and the following useful formulas:
\begin{equation}
\begin{aligned}
\{\gamma_5,\gamma^\mu\}\ =\ 0~,~~~ [\gamma_5,\gamma^{\mu\nu}]&\ =\ 0~,~~~\gamma^{\mu\nu}\ =\ -\tfrac{1}{2}\eps^{\mu\nu\rho\sigma}\gamma_5\gamma_{\rho\sigma}~,\\
{}[\gamma_5,\gamma^\mu]&\ =\ -\tfrac{1}{3}\eps^{\mu\nu\kappa\lambda}\gamma_\nu\gamma_\kappa\gamma_\lambda~,\\
 \{\gamma^\nu,\gamma^{\rho\sigma}\}\ =\ 2\eps^{\nu\rho\sigma\kappa}\gamma_\kappa\gamma_5~,&~~~[\gamma^{\mu\nu},\gamma^\rho]\ =\ 2(\delta^{\nu\rho}\gamma^\mu-\delta^{\mu\rho}\gamma^\nu)~,\\
\{\gamma^{\mu\nu},\gamma^{\sigma\rho}\}\ =\ 2\eps^{\mu\nu\sigma\rho}\gamma_5&-2(\delta^{\mu\rho}\delta^{\nu\sigma}-\delta^{\mu\sigma}\delta^{\nu\rho})\unit~.
\end{aligned}
\end{equation}
An explicit embedding of $\sSU(2)$ is given by 
\begin{equation}
 [\gamma^i,\gamma^j]\ =\ 2\eps^{ijk}\gamma_5\gamma^4\gamma_k~,~~~{i,j,k=1,2,3}~.
\end{equation}
The full Lorentz algebra reads as usual:
\begin{equation}\label{id-lorentz-alg}
{}[\gamma^{\mu\nu},\gamma^{\sigma\rho}]\ =\ 2(\delta^{\mu\sigma}\gamma^{\nu\rho}+\delta^{\nu\rho}\gamma^{\mu\sigma}-\delta^{\mu\rho}\gamma^{\nu\sigma}-\delta^{\nu\sigma}\gamma^{\mu\rho})~.
\end{equation}
In arbitrary even dimensions, note that we have for multi-indices $A,B$:
\begin{equation}
 [\gamma^A,\gamma^B]\ =\ 0~,~~~(\{\gamma^A,\gamma^B\}\ =\ 0~,)
\end{equation}
iff $A$ and $B$ have an odd (even) number of common indices.


\begin{thebibliography}{10}

\bibitem{Basu:2004ed}
A.~Basu and J.~A.~Harvey,
{\em The M2-M5 brane system and a generalized Nahm's equation,}
Nucl. Phys. B {\bf 713} (2005)  136 [{\tt hep-th/0412310}].

\bibitem{Bagger:2006sk}
J.~Bagger and N.~Lambert,
{\em Modeling multiple M2's,}
Phys. Rev. D {\bf 75} (2007)  045020 [{\tt hep-th/0611108}].

\bibitem{Bagger:2007jr}
J.~Bagger and N.~Lambert,
{\em Gauge symmetry and supersymmetry of multiple M2-branes,}
Phys. Rev. D {\bf 77} (2008)  065008 [{\tt 0711.0955 [hep-th]}].

\bibitem{Bagger:2007vi}
J.~Bagger and N.~Lambert,
{\em Comments on multiple M2-branes,}
JHEP {\bf 02} (2008)  105 [{\tt 0712.3738 [hep-th]}].

\bibitem{Gustavsson:2007vu}
A.~Gustavsson,
{\em Algebraic structures on parallel M2-branes,}
{\tt 0709.1260 [hep-th]}.

\bibitem{Gustavsson:2008dy}
A.~Gustavsson,
{\em {Selfdual strings and loop space Nahm equations},}
JHEP {\bf 04} (2008)  083 [{\tt 0802.3456 [hep-th]}].

\bibitem{Schwarz:2004yj}
J.~H.~Schwarz,
{\em Superconformal Chern-Simons theories,}
JHEP {\bf 11} (2004)  078 [{\tt hep-th/0411077}].

\bibitem{Mukhi:2008ux}
S.~Mukhi and C.~Papageorgakis,
{\em {M2 to D2},}
JHEP {\bf 05} (2008)  085 [{\tt 0803.3218 [hep-th]}].

\bibitem{Distler:2008mk}
J.~Distler, S.~Mukhi, C.~Papageorgakis, and M.~Van~Raamsdonk,
{\em {M2-branes on M-folds},}
JHEP {\bf 05} (2008)  038 [{\tt 0804.1256 [hep-th]}].

\bibitem{Banerjee:2008pd}
S.~Banerjee and A.~Sen,
{\em {Interpreting the M2-brane action},}
{\tt 0805.3930 [hep-th]}.

\bibitem{Filippov:1985aa}
V.~T.~Filippov,
{\em $n$-Lie algebras,}
Sib. Mat. Zh. {\bf 26} (1985)  126.

\bibitem{Nagy:2007aa}
P.-A.~Nagy, {\em Prolongations of Lie algebras and applications,} {\tt 0712.1398 [hep-th]}.

\bibitem{Papadopoulos:2008sk}
G.~Papadopoulos,
{\em {M2-branes, 3-Lie Algebras and Pl{\"u}cker relations},}
JHEP {\bf 05} (2008)  054 [{\tt 0804.2662 [hep-th]}].

\bibitem{Gauntlett:2008uf}
J.~P.~Gauntlett and J.~B.~Gutowski,
{\em {Constraining maximally supersymmetric membrane actions},}
{\tt 0804.3078 [hep-th]}.

\bibitem{Ho:2008bn}
P.-M.~Ho, R.-C.~Hou, and Y.~Matsuo,
{\em {Lie 3-Algebra and multiple M2-branes},}
JHEP {\bf 06} (2008)  020 [{\tt 0804.2110 [hep-th]}].

\bibitem{Papadopoulos:2008gh}
G.~Papadopoulos,
{\em {On the structure of k-Lie algebras},}
Class. Quant. Grav. {\bf 25} (2008)  142002 [{\tt 0804.3567 [hep-th]}].

\bibitem{Bandres:2008kj}
M.~A.~Bandres, A.~E.~Lipstein, and J.~H.~Schwarz,
{\em {Ghost-free superconformal action for multiple M2-branes},}
JHEP {\bf 07} (2008)  117 [{\tt 0806.0054 [hep-th]}].

\bibitem{Benna:2008zy}
M.~Benna, I.~Klebanov, T.~Klose, and M.~Smedback,
{\em {Superconformal Chern-Simons theories and $AdS_4/CFT_3$ correspondence},}
{\tt 0806.1519 [hep-th]}.

\bibitem{Mauri:2008ai}
A.~Mauri and A.~C.~Petkou,
{\em {An N=1 superfield action for M2 branes},}
{\tt 0806.2270 [hep-th]}.

\bibitem{Dzhuma:2002aa}
A.~S.~Dzhumadil'daev,
{\em Representations of $n$-Lie algebras,}
{\tt math.RT/0202041}.

\bibitem{Kasymov1983:aa}
S.~M.~Kasymov,
{\em Theory of n-Lie algebras,}
Algebra i Logika {\bf 26} (1983)  277.

\bibitem{Polchinski:1998rr}
J.~Polchinski,
{\em String theory. Vol. 2: Superstring theory and beyond,}
Cambridge, UK: Univ. Pr. (1998).

\bibitem{Gran:2008vi}
U.~Gran, B.~E.~W.~Nilsson, and C.~Petersson,
{\em {On relating multiple M2 and D2-branes},}
{\tt 0804.1784 [hep-th]}.

\bibitem{Gomis:2008uv}
J.~Gomis, G.~Milanesi, and J.~G.~Russo,
{\em {Bagger-Lambert theory for general Lie algebras},}
JHEP {\bf 06} (2008)  075 [{\tt 0805.1012 [hep-th]}].

\bibitem{Benvenuti:2008bt}
S.~Benvenuti, D.~Rodriguez-Gomez, E.~Tonni, and H.~Verlinde,
{\em {N=8 superconformal gauge theories and M2 branes},}
{\tt 0805.1087 [hep-th]}.

\bibitem{Ho:2008ei}
P.-M.~Ho, Y.~Imamura, and Y.~Matsuo,
{\em {M2 to D2 revisited},}
JHEP {\bf 07} (2008) 003 [{\tt 0805.1202 [hep-th]}].

\bibitem{Lin:2008qp}
H.~Lin, {\em Kac-Moody extensions of 3-algebras and M2-branes,} 
JHEP {\bf 07} (2008) 136 [{\tt 0805.4003 [hep-th]}].

\bibitem{FigueroaO'Farrill:2008zm}
J.~Figueroa-O'Farrill, P.~de~Medeiros, and E.~Mendez-Escobar,
{\em {Lorentzian Lie 3-algebras and their Bagger-Lambert moduli space},}
JHEP {\bf 07} (2008) 111 [{\tt 0805.4363 [hep-th]}].

\bibitem{Gomis:2008be}
J.~Gomis, D.~Rodriguez-Gomez, M.~Van~Raamsdonk, and H.~Verlinde,
{\em {The superconformal gauge theory on M2-branes},}
{\tt 0806.0738 [hep-th]}.

\bibitem{Figueroa-O'Farrill:arXiv0805.4760}
J.~Figueroa-O'Farrill,
{\em Lorentzian Lie $n$-algebras,}
{\tt 0805.4760 [math.RT]}.

\bibitem{deMedeiros:2008bf}
P.~de~Medeiros, J.~Figueroa-O'Farrill, and E.~Mendez-Escobar,
{\em {Metric Lie 3-algebras in Bagger-Lambert theory},}
{\tt 0806.3242 [hep-th]}.

\bibitem{Aharony:2008ug}
O.~Aharony, O.~Bergman, D.~L.~Jafferis, and J.~Maldacena,
{\em {N=6 superconformal Chern-Simons-matter theories, M2-branes and their
  gravity duals},}
{\tt 0806.1218 [hep-th]}.

\bibitem{Minahan:2008hf}
J.~A.~Minahan and K.~Zarembo,
{\em {The Bethe ansatz for superconformal Chern-Simons},}
{\tt 0806.3951 [hep-th]}.

\bibitem{Wess:1992cp}
J.~Wess and J.~Bagger,
{\em Supersymmetry and supergravity,}
Princeton, USA: Univ. Pr. (1992).

\bibitem{Zupnik:1988en}
B.~M.~Zupnik and D.~G.~Pak,
{\em {Superfield formulation of the simplest three-dimensional gauge theories
  and conformal supergravities},}
Theor. Math. Phys. {\bf 77} (1988)  1070.

\bibitem{Ivanov:1991fn}
E.~A.~Ivanov,
{\em Chern-Simons matter systems with manifest N=2 supersymmetry,}
Phys. Lett. B {\bf 268} (1991)  203.

\bibitem{Lister:1952aa}
W.~G.~Lister,
{\em A structure theory of Lie triple systems,}
Trans. Am. Math. Soc. {\bf 72} (1952)  217.

\bibitem{Faulkner:1980aa}
J.~R.~Faulkner, {\em Dynkin diagrams for Lie triple systems,} J. of Algebra {\bf 62} (1980) 384.

\bibitem{Benito:1999aa}
P.~Benito, C.~Draper, and A.~Elduque,
{\em On some algebras related to simple Lie triple systems,}
J. of Algebra {\bf 219} (1999)  234.

\bibitem{Gustavsson:2008bf}
A.~Gustavsson,
{\em {One-loop corrections to Bagger-Lambert theory},}
{\tt 0805.4443 [hep-th]}.

\bibitem{Bagger:2008se}
J.~Bagger and N.~Lambert,
{\em {Three-algebras and N=6 Chern-Simons gauge theories},}
{\tt 0807.0163 [hep-th]}.

\end{thebibliography}
\end{document}